\documentclass[a4paper, amsfonts, amssymb, amsmath, reprint, showkeys, nofootinbib,superscriptaddress,longbibliography]{revtex4-2}

\usepackage[left=23mm,right=13mm,top=35mm,columnsep=15pt]{geometry} 

\usepackage{mhchem}
\usepackage{braket}
\usepackage{gensymb}
\usepackage{graphicx}

\usepackage[english]{babel}
\usepackage[utf8]{inputenc}

\usepackage{xurl}
\usepackage[pdftex, pdftitle={Article}, pdfauthor={Author}]{hyperref} 

\begin{document}

\title{Wide-field mid- to long-wave infrared imaging with undetected photons}
\author{Vladimir Kornienko}
\email{v.kornienko@imperial.ac.uk}
\affiliation{Department of Physics, Imperial College London, Prince Consort Rd, London SW7 2BW, United Kingdom}
\author{Nathan Gemmell}
\affiliation{Department of Physics, Imperial College London, Prince Consort Rd, London SW7 2BW, United Kingdom}
\author{Caiyi Liu}
\affiliation{Department of Physics, Imperial College London, Prince Consort Rd, London SW7 2BW, United Kingdom}
\author{Asteria Chen}
\affiliation{Department of Physics, Imperial College London, Prince Consort Rd, London SW7 2BW, United Kingdom}
\author{Chris Phillips}
\affiliation{Department of Physics, Imperial College London, Prince Consort Rd, London SW7 2BW, United Kingdom}
\author{Rupert Oulton}
\affiliation{Department of Physics, Imperial College London, Prince Consort Rd, London SW7 2BW, United Kingdom}

\begin{abstract}
Quantum imaging with undetected photons (QIUP) allows an object to be probed at mid-infrared frequencies by only measuring interference in the visible range, thus leveraging silicon camera technology. We show that non-collinear phase-matching in a silver thiogallate (\ce{AgGaS2}) crystal enables wide-field QIUP in the wavelength range of $6$ -- $10$ $\mu$m (1670 -- 1000~cm$^{-1}$). A combination of coherent detection and infrared photons being ``undetected'' enables imaging at ${\sim}100$ times better than the background-limited infrared photodetection (BLIP) limit. At 8~$\mu$m, our images have over $8000\pm100$ resolvable elements with a $297 \pm 5 \, \mu$m resolution, and 10~s acquisition time. Our results pave the way to fast, background-noise-free, room-temperature, spectrally-selective mid-infrared imaging.
\end{abstract}

\maketitle

\section{Introduction}

Imaging outside the visible spectrum enables contrast mechanisms that reveal structure, function, and composition beyond our direct perception. Among these, infrared (IR) spectroscopy is particularly powerful because it identifies the spectral signatures of molecular functional groups; such ``fingerprint'' absorption bands occur at roughly the same wavelength independent of the group's local surrounding environment \cite{2020_Bec_IR_applications}. This intrinsic specificity makes mid- to long-wave IR imaging uniquely attractive for label-free analysis in biology, materials science, and chemical sensing. While the motivation is clear, several technical challenges make spectroscopic IR imaging technically difficult and often impractically slow \cite{2007_Piotrowski_BLIP}. Firstly, thermal background radiation generates fluctuations which, at room temperature, peak at around 10~$\mu$m wavelength (${\sim}1000$~cm$^{-1}$), 
setting the minimum detectable signal level. In this background-limited infrared photo-detection (BLIP) regime, the attainable signal-to-noise ratio is limited by the shot noise associated with the environmental thermal background. Furthermore, large background thermal flux from a scene can saturate even cooled IR detectors, imposing stringent dynamic-range requirements and often necessitating narrowband spectral filtering to suppress out-of-band radiation. Finally, IR cameras rely on specialized detector materials, fabrication processes, and cryogenic operation, and generally offer fewer pixels with lower frame rates than their technically honed visible counterparts.

Indirect IR detection can be realized with nonlinear interferometry (NLI)\cite{2016_Chekhova_Ou_NLI}, which relies on the intrinsic indistinguishability of photon paths as observed in early induced coherence experiments \cite{1991_Mandel}. The underlying principle dates back to Ramsey’s method of separated fields \cite{1950_Ramsey}. In 2014, NLI-based imaging was demonstrated and the term quantum imaging with undetected photons (QIUP) was coined \cite{2014_Zeilinger_QIUP}. In QIUP, an object placed in the `idler' ($\omega_i$) beam --- IR in this Letter --- modifies the interference pattern of the visible (`signal', $\omega_s$) beam. Because the IR light is not detected, the measurement is immune to background thermal radiation \cite{2023_Nathan_RemoveThermalBG}, resulting in faster, noise-free imaging. The unique combination of two key features ---  coherent detection and infrared photons being ``undetected'' --- is a key advantage of QIUP over up-conversion imaging \cite{2022_UpConversion_Imaging}, homodyne detection \cite{2023_Homodyne}, and so-called ``ghost imaging'' \cite{2010_Shapiro_GhostImaging,2015_NonDegenerate_GI}.

Several methods of extending single-mode IR sensing with undetected photons up to 11~$\mu$m wavelength for spectroscopy applications have been demonstrated \cite{2016_Krivitsky_IRspectrWvisLight,2025_Pertsch,2025_Clark_GasSensing_3um2}. In the optical domain, the longest IR wavelengths where multi-mode QIUP has been demonstrated to date are 4.3~$\mu$m in pp-KTP \cite{2020_Ramelow_Fernsehturm,2025_Chekhova_HighGain_MidIR_3um,2025_Graefe}, and 3.4~$\mu$m \cite{2025_Paterova_Microfluidics} in PPLN.
QIUP has also been demonstrated \cite{2025_Kutas_THz_NLI} in the terahertz-wave range; however, implementations in the mid- to long-infrared fingerprint region have not yet been reported. A major difficulty when switching from a sensing approach to imaging is the lack of easily accessible components that span all three spectral ranges (pump, signal, and idler). While the properties of the nonlinear crystal are paramount, high transmittance, low fluorescence, and low chromatic dispersion are essential in all the elements of the system, and this becomes increasingly challenging over larger spectral ranges. 
Hence, alternative system designs are desirable, especially new approaches to partition the three interacting waves.

\begin{figure}[htp]
\centering
\fbox{\includegraphics[width=\linewidth]{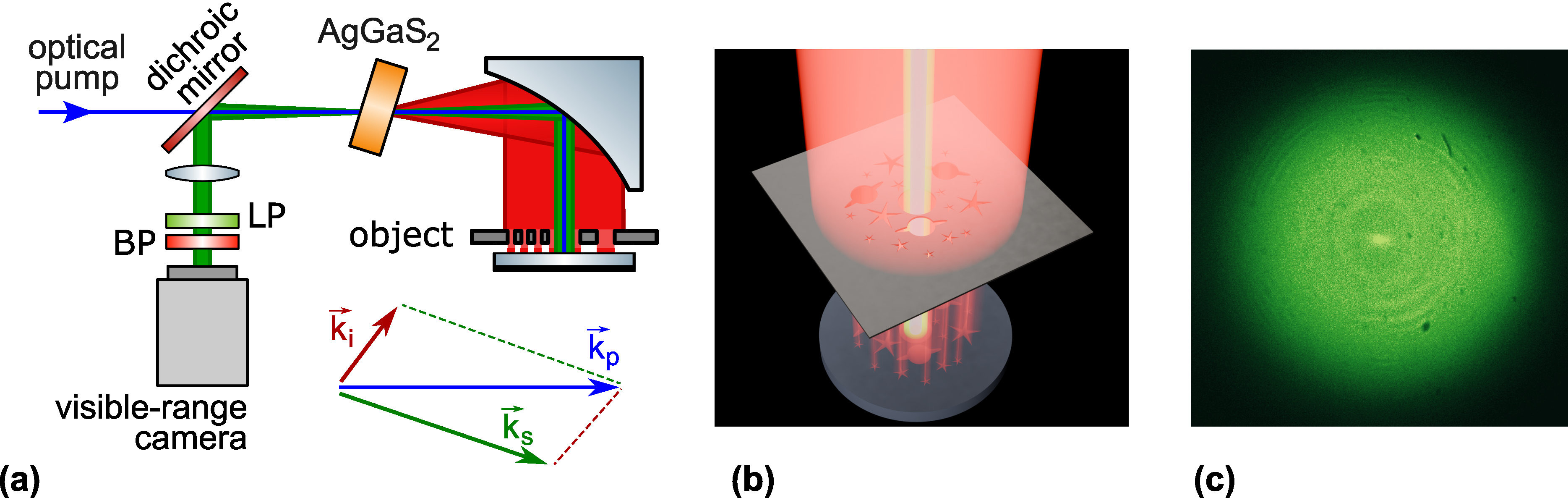}}
\caption{\textbf{(a)} Nonlinear interferometry set-up featuring non-collinear phase matching, folded geometry and common path arrangement. \textbf{(b)} Schematic of an object (shadow mask) with a hole in the center to let the pump and signal beams through. The outer annulus of the idler beam can be used for imaging. \textbf{(c)} Raw camera image showing interference only in the regions not blocked by the shadow mask. Intensity is shown in linear contrast scale.}
\label{Fig:ExpSetUp_partA}
\end{figure}

Prospective nonlinear materials for QIUP are the same as those used in sum- and difference-frequency generation in mid- to long-wave IR range, and include \ce{AgGaS2}, \ce{AgGaSe}, \ce{BGGS}, \ce{BGGSe} \cite{Dmitriev_HandbookOfNLcryst}. \ce{AgGaS2} has received the most attention recently, with spectroscopy performance demonstrated up to 11~$\mu$m~\cite{2022_Paterova_OffAxisParabola,2022_Mukai_AGS_Spectroscopy} for external samples and up to 21~$\mu$m for studying the nonlinear crystal's intrinsic absorption \cite{2026_Paterova_UpTo21um}. However, it is not clear if wide-field imaging at such large wavelengths is feasible since the number of modes used for imaging is expected to drop with wavelength \cite{2022_Ramelow_position_corr}. To date, no wide-field QIUP beyond 4.3~$\mu$m has been reported.

Here, we report a scheme for wide-field imaging at $6$ -- $10$ $\mu$m (1670 -- 1000~cm$^{-1}$) wavelengths based on nonlinear interferometry in a silver thiogallate (silver gallium sulfide, \ce{AgGaS_2}, AGS) crystal with non-collinear phase matching.
We find that the resolution is uniform across the field of view and the total number of resolvable elements in the image exceeds $8000\pm100$.
We also provide a comparison of two methods for processing interferometric data to produce images; one based on collecting a series of images at different interferometric phase values, and a single-frame acquisition approach based on off-axis holography.

\section{Experimental set-up}

Our experimental set-up is shown in Fig.~\ref{Fig:ExpSetUp_partA}a. We used 50~mW of continuous-wave radiation ($\lambda_p = 700$--$740\textrm{~nm}$) from a Ti:Sapphire laser (\textit{Spectra-Physics Matisse}) 
to optically pump the nonlinear crystal (\ce{AgGaS2}). As a result of the spontaneous parametric down-conversion (SPDC),
a pair of a signal ($\lambda_s$) and an idler ($\lambda_i$) photons are generated, where the frequencies of interacting waves satisfy the energy conservation law: $\omega_p = \omega_s + \omega_i$ ($\lambda_p^{-1} = \lambda_s^{-1} + \lambda_i^{-1}$). The SPDC intensity is maximized when the phase-matching condition is met: $\vec{k}_p = \vec{k}_s + \vec{k}_i$, as illustrated by the inset to Fig.~\ref{Fig:ExpSetUp_partA}a. An off-axis parabolic mirror is used to collimate all three beams while avoiding chromatic aberration. The idler beam is used to probe the object to be imaged and is never itself detected directly, while the signal beam is registered with a CMOS camera (\textit{Hamamatsu ORCA-Quest}), and its interference pattern is analyzed. A folded geometry 
allows us to use a single crystal to establish nonlinear interference: all three 
beams pass back through the crystal. On the second pass, the pump is able to generate new signal and idler photons. When the system is aligned to ensure a high degree of indistinguishability between photon pairs generated on the two distinct passes of the crystal, nonlinear interference is established. A conventional dichroic mirror is used to separate the signal and pump waves. 
Since this is effectively a common path arrangement in terms of opto-mechanics, we scan the interferometric phase $\phi$ by displacing not the end mirror of the interferometer but the crystal from the focal point of the parabolic mirror. In principle, the three beams could be spatially separated using a suitable mirror configuration.


The idler wavelength $\lambda_i$ can be tuned by changing the bandpass filter in front of the CMOS camera, the \ce{AgGaS2} crystal orientation, or the pump wavelength. At least two of these need to be varied to adjust both the field of view and the imaging wavelength simultaneously. In our experiment, we fixed a 10~nm bandwidth $\lambda_s$ bandpass filter at 800~nm central wavelength and varied the pump wavelength $\lambda_p$ from 700~nm to 740~nm and the crystal orientation with rotation angles of $\sim \pm 10\degree$. This allowed us to perform imaging in the 6--10 $\mu$m range with an equivalent idler bandwidth of around 100~nm. An object was placed at the backward focal plane of the parabolic mirror, close to the surface of the flat mirror, often referred to as the far-field imaging configuration or Fourier plane. This mapped the object transmittance to the wavevectors of the idler beam, thus using momentum correlations for image reconstruction \cite{1998_TransferOfAngularMomentum}.

Although the signal $(\lambda_s)$ and the idler $(\lambda_i)$ beams follow a common path arrangement, we use a large $\left( \approx \lambda_i / \lambda_s \sim 10 \textrm{~times}\right)$ difference in beam diameters to enable their separation. Consequently, the small central part of the mask cannot be used for imaging (see Fig.~\ref{Fig:ExpSetUp_partA}b), but we can use the outer annulus of the idler beam that does not overlap with the signal. 
The object transmittance at the idler wavelength is reconstructed from the visibility and phase of the interference pattern (Fig.~\ref{Fig:ExpSetUp_partA}c) in the signal beam recorded by the camera. Both amplitude and phase transmittance data are retrieved. 

\section{Results \& Discussion}

\begin{figure*}[t!]
\centering
\fbox{\includegraphics[height=3.6cm]{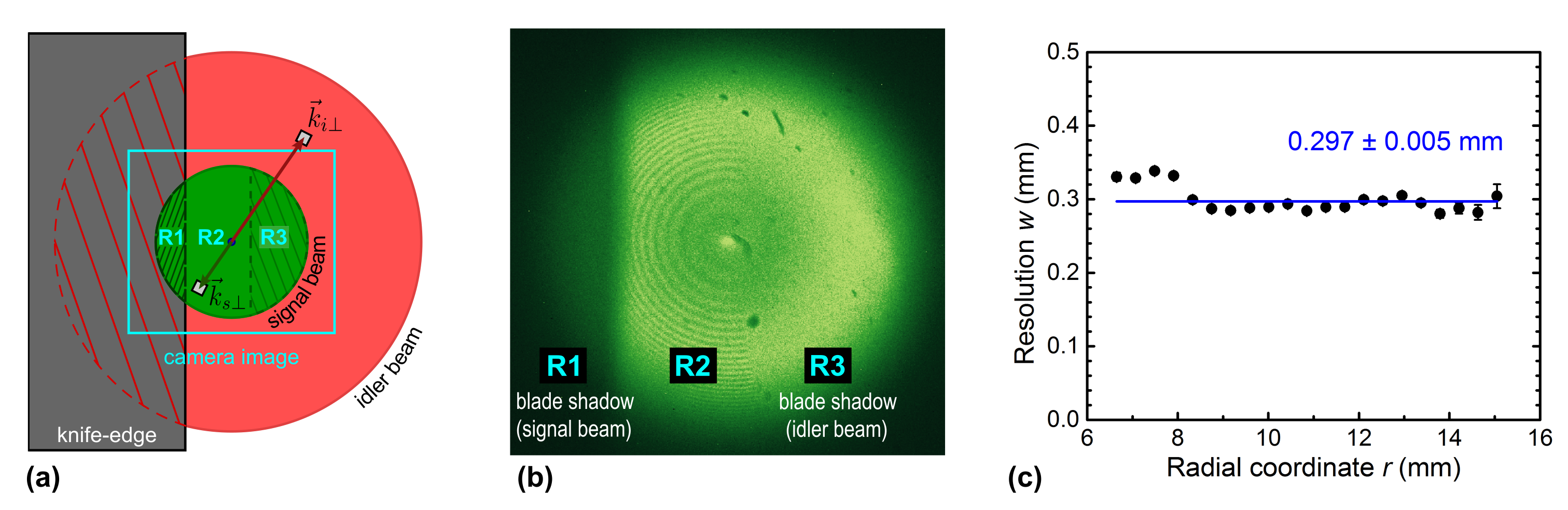}}
\caption{Estimating the resolution of the QIUP system with a knife-edge test. The idler beam is ${\approx}\lambda_i / \lambda_s {\sim} 10$ times bigger than the signal beam. \textbf{(a,b)}.  Bringing the knife edge close to the optical axis of the system results in the blade shadow appearing in both signal and idler beams. Three regions in the image can be observed: R$_1$ -- signal beam is blocked, intensity is halved, no interference observed; R$_2$ -- all three beams are intact, nonlinear interference is observed; R$_3$ -- blocking the idler in R$_1$ leads to distinguishability in the photon pairs, thus no interference is observed in the signal beam, but the signal beam intensity is unaffected. \textbf{(c)} Resolution $w$ of the scheme does not depend on the radial coordinate $r$.}
\label{Fig:Resolution}
\end{figure*}

In a nonlinear interference experiment sketched in Fig.~\ref{Fig:ExpSetUp_partA}a, the visible signal beam intensity on the camera has the following dependence \cite{2022_Lemos_Tutorial} on the object's transmittance $\tau \equiv \lvert \tau\rvert \exp({i \phi_\textrm{obj})}$ at the IR idler frequency:
\begin{equation}
I_s \propto 1 + {\lvert \tau \rvert}^2 \cos(\phi + 2\phi_\textrm{obj}) \; ,
\label{Eq:CameraIntensity}
\end{equation}
where $\phi$ is the interferometric phase governed by the interferometer configuration, and the idler beam passes the object twice. Taking a series of camera images at different values of $\phi$ allows the visibility $V \equiv (I_\textrm{MAX}-I_\textrm{MIN})/(I_\textrm{MAX}+I_\textrm{MIN})$ to be extracted at each pixel of the visible-range CMOS camera. One can then reconstruct the object's complex transmittance coefficient from Eq.~\ref{Eq:CameraIntensity}, including the phase $(\phi_\textrm{obj})$. For example, a power transmittance $T \equiv {\lvert \tau \rvert}^2$ of the object at the idler wavelength in our case equals the visibility of the interference pattern:${\lvert \tau \rvert}^2 \equiv T = \textrm{Vis}$. When the interference fringes are sufficiently dense, the visibility can also be extracted from a single image via a spatial Fourier transform, as shown in Refs.~\cite{2023_Pearce_Practical_QIUP} and \cite{2024_Pearce_TiltedFringes}.

The resolution was estimated with a knife-edge test (Fig.~\ref{Fig:Resolution}). A razor blade was put in place of the object in Fig.~\ref{Fig:ExpSetUp_partA}a.
Translating the blade allowed us to block a variable fraction of the idler beam, while leaving the pump and signal beams unaffected.
Blocking the idler beam makes the two passes through the crystal distinguishable and hence destroys the interference in the region (R$_3$). The resulting visibility curves were fitted along the concentric arcs of radii $r$ with an error function: $V(x) = V_0 + A \, \textrm{erf} ( (x-x_0)\sqrt{2}/w ) $, to extract the resolution $w$ of the scheme as a function of $r$ (Fig.~\ref{Fig:Resolution}c). At an 8~$\mu$m imaging wavelength, the resolution was found to be $w=297\pm5\,\mu$m. Given that the field of view is ${{\sim}\pi (30\textrm{~mm}/2)^2}$, we thus estimate the total number of resolvable elements to be $8000\pm100$, equivalent to the number of modes of the SPDC joint angular spectrum.
Notably, the resolution was the same across the whole image plane, and was not affected by the interference fringes being denser at the outer edge of the image, as long as the wavelength range used for the imaging was itself sufficiently narrow-band.

When the blade was moved even closer to the system's optical axis, its shadow also appeared in the camera (Fig.~\ref{Fig:Resolution}b). Blocking the signal SPDC beam from the first pass of the pump beam through the crystal leaves only the signal SPDC from the second pass, and so the corresponding part (R$_1$) of the image appears darker on the camera. The blade shadows (R$_1$) and ({R$_3$}) for the signal and idler beams respectively appear on opposite sides of the image since the transverse wavevectors $\vec{k}_{s \perp}$ and $\vec{k}_{i \perp}$ follow momentum conservation, $\vec{k}_{s \perp} +\vec{k}_{i \perp}=0$.


\begin{figure}[bp]
\centering
\fbox{\includegraphics[width=0.8 \linewidth]{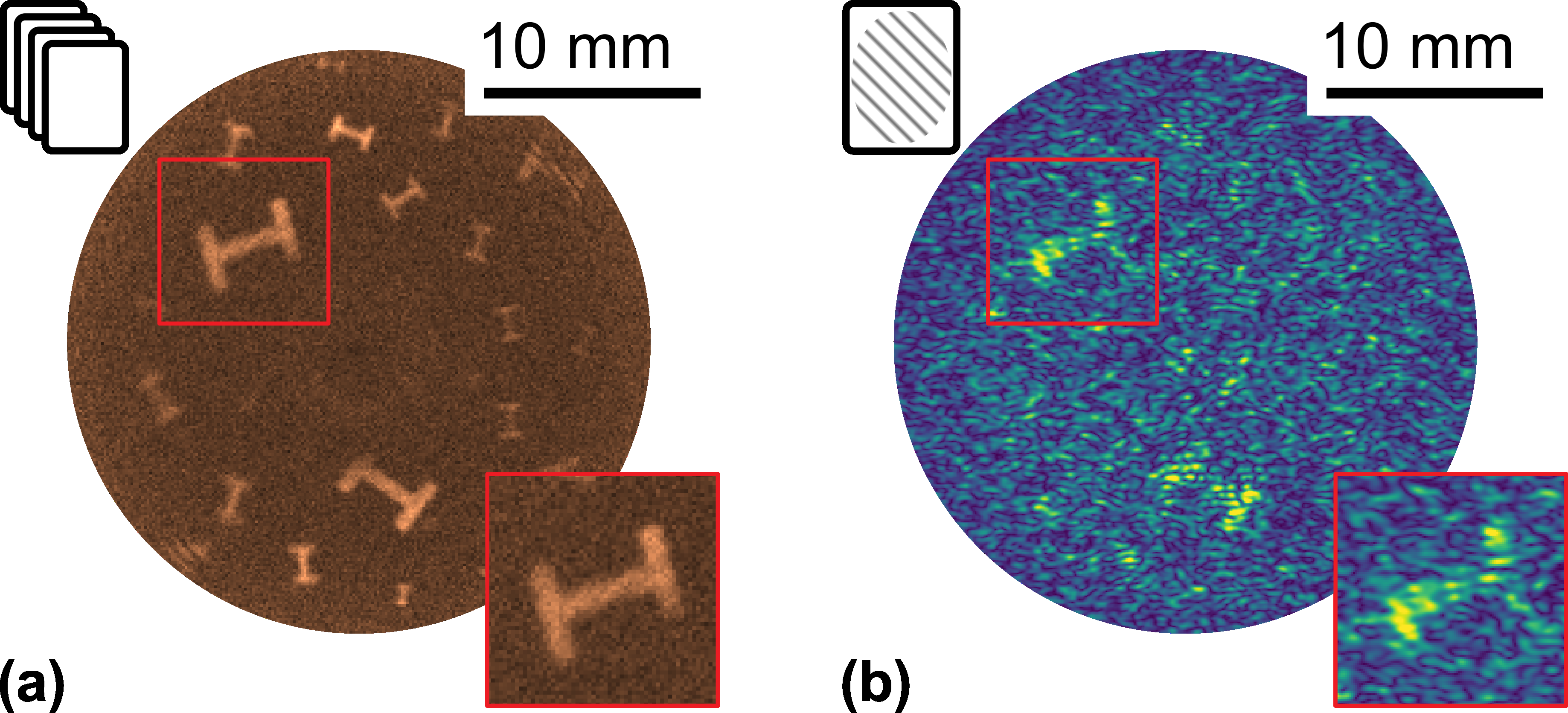}}
\caption{Interference pattern visibility at 8~$\mu$m idler wavelength computed from: \textbf{(a)}~a~dataset of 180 files corresponding to different interferometric phase values $\phi$, or \textbf{(b)}~a single camera image using the off-axis holography approach \cite{2024_Pearce_TiltedFringes}. One image acquisition time was 10~s. Scale bar shows the physical size at the imaging plane.}
\label{Fig:ExpSetUp_partB}
\end{figure}

\begin{figure*}[pt!]
\centering
\fbox{\includegraphics[height=3.6cm]{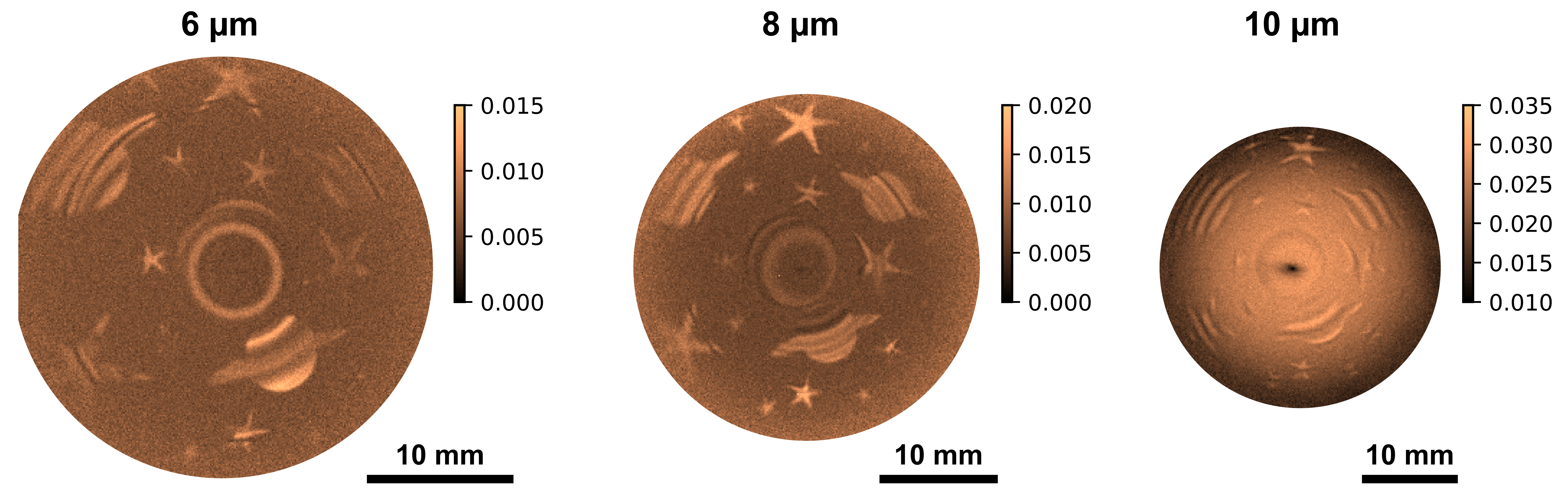}}
\caption{Computed interference pattern visibility at idler wavelengths of 6~$\mu$m, 8~$\mu$m,
and 10~$\mu$m. Scale bar shows the physical size at the imaging plane. Original datasets contain 180 images of 10~s acquisition time each.}
\label{Fig:Imaging}
\end{figure*}

Examples of reconstructed images of shadow masks are shown in Fig.~\ref{Fig:ExpSetUp_partB}.
The masks were cut in a thin (0.3~mm) metal foil, and acted as $T=0$ or $1$ transmittance object with no additional phase shift. As illustrated in Fig.~\ref{Fig:ExpSetUp_partA}b, pump and signal beams passed unaffected through the central hole in the mask only, 
and the central region of the idler beam was not suitable for imaging. The mid-infrared idler beam was used to create an image of the outer part of the mask. The signal beam never passed through the outer annulus of the shadow mask. Images in Fig.~\ref{Fig:ExpSetUp_partB} show the interference fringes visibility plotted in linearly scaled color. Openings in the mask result in high interference visibility, and are seen as bright regions in Fig.~\ref{Fig:ExpSetUp_partB}. In other regions, the mask blocks the idler wave thus destroying the interference, which corresponds to a darker color. Detailed multi-file scans involved up to 180 frames with 10 second acquisition time each.

If the interference fringes are sufficiently dense compared with the object's dimensions, one can implement the imaging using an off-axis holography approach \cite{2024_Pearce_TiltedFringes}, where a single image is used to reconstruct the object's transmittance. The results of processing a single image from the dataset used to reconstruct image Fig.~\ref{Fig:ExpSetUp_partB}a is shown in panel (b). This single-frame approach offers a significant speedup, bringing the total time from 30 minutes down to 10 seconds, albeit at the cost of reduced image quality. This presents a significant speed-up compared to scanning Fourier Transform IR (FTIR) Spectroscopy, where acquiring an image with a similar resolution and field of view takes approximately 6~hours.

The images shown above, and especially the single-frame image in Fig.~\ref{Fig:ExpSetUp_partB}, contain artifacts that arise chiefly from the nonlinear dependence of the interferometric phase $\phi$ and of the imaging wavelength $\lambda_i$ on the coordinate in the image. As discussed in \cite{2022_Paterova_OffAxisParabola}, displacing the crystal by distance $\epsilon$ from the focal point of the parabolic mirror changes the interferometric phase in Eq.~\ref{Eq:CameraIntensity} as follows:
\begin{equation}
\phi = \Delta k_{\parallel} L_{cr} / 2 + \Delta k_{\parallel}^\prime \epsilon /2 \, ,
\end{equation}
where $L_{cr}$ is the nonlinear crystal length, $\Delta k_{\parallel}$ and $\Delta k_{\parallel}^\prime$ are the longitudinal phase mismatch values inside and outside the crystal, respectively. The interference fringes are larger in the middle and denser at the outer edge of the image. We scanned the interferometric phase $\phi$ by translating the crystal by about 1~mm with a 5~$\mu$m step. The number of interference fringes observed was also changing. Interferometric phase $\phi$ is linear in crystal displacement $\epsilon$, but is not linear in the radial coordinate across the image, as $\Delta k_\parallel^\prime$ can be approximated as scaling quadratically with the coordinate $r$ from the center of the image.
Hence, crystal displacement $\epsilon$ changes the period of the interference pattern, and visibility data reconstruction requires advanced data processing. For example, if Fourier-transform-based processing similar to Ref.~\cite{2023_Pearce_Practical_QIUP} is used, different Fourier components correspond to different annuli of the SPDC ring.

Finally, we demonstrate that our system is capable of wide-field imaging in the important ``fingerprint'' range from 6~$\mu$m to 10~$\mu$m. Images of a shadow mask cut in a thin (0.3~mm) metal foil for three wavelengths (6, 8, and 10~$\mu$m) are shown in Fig.~\ref{Fig:Imaging}. The imaging wavelength $\lambda_i$ was tuned by changing the pump wavelength $\lambda_p$ and rotating the crystal, while the shadow mask position was kept fixed. Distortions in the reconstructed visibility maps 
are mostly due to the crystal holder frame blocking a portion of the idler beam and changes in the SPDC intensity distribution across the image, as well as the data processing issues discussed above.

The total number of signal photons detected within the field of view is around $10^7$ per second. As the signal and the idler photons are created in pairs under the SPDC process, we can estimate the overall power in the idler channel to be 0.2~pW ($2\times 10^{-13}$~W), corresponding to $7\times10^{-18}$~W per camera pixel. The noise equivalent power~(NEP) of a BLIP-limited photodetector with similar parameters (100~nm idler bandwidth, 8~$\mu$m central wavelength, 7{\degree} field of view) and the size of one camera pixel ($8.6\times8.6$~$\mu$m$^{2}$ area) is ${\sim}2\times10^{-15}$~W~Hz$^{-1/2}$
 \cite{2007_Piotrowski_BLIP}, and the total blackbody radiation flux density is ${\sim}2.1\times10^{-10}$~W. This illustrates that quantum imaging with undetected photons can detect infrared light at levels that are at least 2 orders of magnitude less than the respective BLIP limit. Combined with acquisition times in the range of minutes and spectral selectivity, this makes QIUP a promising technique for mid- to long-wave IR spectroscopy and imaging.


\section{Summary and Outlook}

We have shown that non-collinear phase-matching in a silver thiogallate (\ce{AgGaS2}) crystal enables wide-field imaging in the range of $6$ -- $10$ $\mu$m (1670 -- 1000~cm$^{-1}$) based on nonlinear interferometry (NLI), also referred to as quantum imaging with undetected photons (QIUP). The total number of resolvable elements in the image exceeds $8000\pm100$ in the field of view. The interferometric phase can be scanned by changing the distance between the nonlinear crystal and the parabolic mirror. Analytical treatment and experimental data show that the resolution is uniform across the field of view. At $8 \, \mu$m imaging wavelength, the resolution is $297 \pm 5 \, \mu$m, and the field of view (defined at 90\% intensity level) is $\pi (30\textrm{mm}/2)^2 \approx 700\textrm{mm}^2$. The imaging system can be operated in two distinct regimes: phase scanning interferometry using multi frame acquisition to produce high quality images; or single-frame off-axis holography for up to $100$ times faster image acquisition rates. The imaging system is resilient to background thermal noise, producing images with IR power approximately 2 orders of magnitude below the background-limited infrared photodetection (BLIP) limit. This is possible, since QIUP does not require direct detection of IR radiation, in contrast to ghost imaging. Our results pave the way to fast, background-noise-free, spectrally-selective infrared imaging with uncooled detectors, for applications in label-free biomedical imaging, materials science, and chemical sensing.

\section{Funding} This work is supported by the UK Quantum Technology Hub in Sensing Imaging and Timing (EP/Z533166/1) and National Institute for Health and Care Research (NIHR517225).

\section{Disclosures} The authors declare no conflicts of interest.

\section{Data Availability Statement} Data underlying the results presented in this paper are not publicly available at this time but may be obtained from the authors upon reasonable request.

\bibliography{AGS_imaging}

\end{document}